\pdfoutput=1
\documentclass[prb,aps,twocolumn,epsfig, doublespace]{revtex4}
\usepackage{graphicx}
\usepackage{dcolumn}
\usepackage{bm}
\makeatletter

\newcommand{\Rmnum}[1]{\expandafter\@slowromancap\romannumeral #1@}
\makeatother
\begin{document}
\title {High contrast imaging and thickness determination of graphene with
in-column secondary electron microscopy}
\author{Vidya Kochat,$^1$ Atindra Nath Pal,$^1$ Sneha E. S.,$^1$ Arjun B. S.,$^2$ Anshita Gairola,$^2$ S. A. Shivashankar,$^2$
Srinivasan Raghavan,$^2$ and Arindam Ghosh$^1$} \vspace{1.5cm}
\address{$^1$ Department of Physics, Indian Institute of Science, Bangalore 560 012, India}
\address{$^2$ Materials Research Center, Indian Institute of Science, Bangalore 560 012, India}

\begin{abstract}

We report a new method for quantitative estimation of graphene layer thicknesses using high contrast imaging of graphene films on insulating substrates with a scanning electron microscope. By detecting the attenuation of
secondary electrons emitted from the substrate with an in-column low-energy
electron detector, we have achieved very high thickness-dependent contrast that
allows quantitative estimation of thickness up to several graphene layers. The
nanometer scale spatial resolution of the electron micrographs also allows a
simple structural characterization scheme for graphene, which has been applied
to identify faults, wrinkles, voids, and patches of multilayer growth in
large-area chemical vapor deposited graphene. We have discussed the factors,
such as differential surface charging and electron beam induced current, that
affect the contrast of graphene images in detail.

\end{abstract}


\maketitle

\section{Introduction}

As a two dimensional crystal of carbon atoms, graphene has attracted wide
attention due to its potential applications in the field of ultra-fast
electronics, sensing and photovoltaics, apart from providing a testing bed of
the new and exotic physical phenomena. It has been observed that the
electrical~\cite{{bilayer},{trilayer}}, thermal~\cite{thermopower}, mechanical~\cite{mechanical} and optical~\cite{optical_1} properties of graphene films depend strongly on the number of graphene layers, because the band structure of
multilayer graphene systems differs radically from monolayer graphene due
to the interlayer coupling~\cite{McCann_bilayer}. The number of layers determines the response of
graphene to external factors too, such as its coupling to underlying
substrates, disorder and adsorbates~\cite{atin APL,{atin PRL},{noise arxive}}. Hence with advances in new techniques for
growing graphene at a macroscopic scale, which includes epitaxial growth on SiC~\cite{SiC}
or chemical vapor deposition on transition metals (copper, nickel, tungsten
etc.)~\cite{{CVD_science},{CVD_nanoletters}}, there is a growing emphasis on a close monitoring of graphene thickness
in a rapid and non-invasive manner with nanometer scale spatial resolution. In
this letter we show that scanning electron microscopy (SEM) with an in-column
secondary electron (SE) detector is an excellent tool for high-contrast
layer-sensitive imaging of graphene, which is simple and overcomes many of the
drawbacks of commonly used methods.

Several techniques have been used to characterize the thickness of graphene
layers. This includes optical microscopy~\cite{{optical_1},{optical_2}}, atomic force microscopy (AFM)~\cite{AFM}, Raman
spectroscopy~\cite{{Raman_1},{Raman_2},{Raman_referee}}, contrast spectroscopy~\cite{contrast spectroscopy}, Rayleigh imaging~\cite{Rayleigh spectroscopy}, low energy electron diffraction, angle-resolved ultraviolet photoemission spectroscopy~\cite{LEED} and
Auger electron spectroscopy~\cite{Auger}. Optical microscopy is limited to identifying
graphene layers on $\sim300$~nm SiO$_2$. Raman spectroscopy is presently the
most common technique used to distinguish between monolayer and bilayer
graphene on SiO$_2$, but not regularly on other substrates like SiC, glass, Cu, Ni due to
strong electronic coupling between graphene and the substrate. Moreover, the
diffraction limit of light imposes a restriction on the spatial resolution of
Raman spectroscopy of graphene, as also in the cases of contrast
spectroscopy and Rayleigh imaging. AFM is a direct technique to count
the number of layers in graphene, even in sub-micron samples, but the tip-sample interaction~\cite{AFM}and the topology of the substrate need to be carefully considered.

\begin{figure}[b]
\begin{center}
\includegraphics [width=1\linewidth]{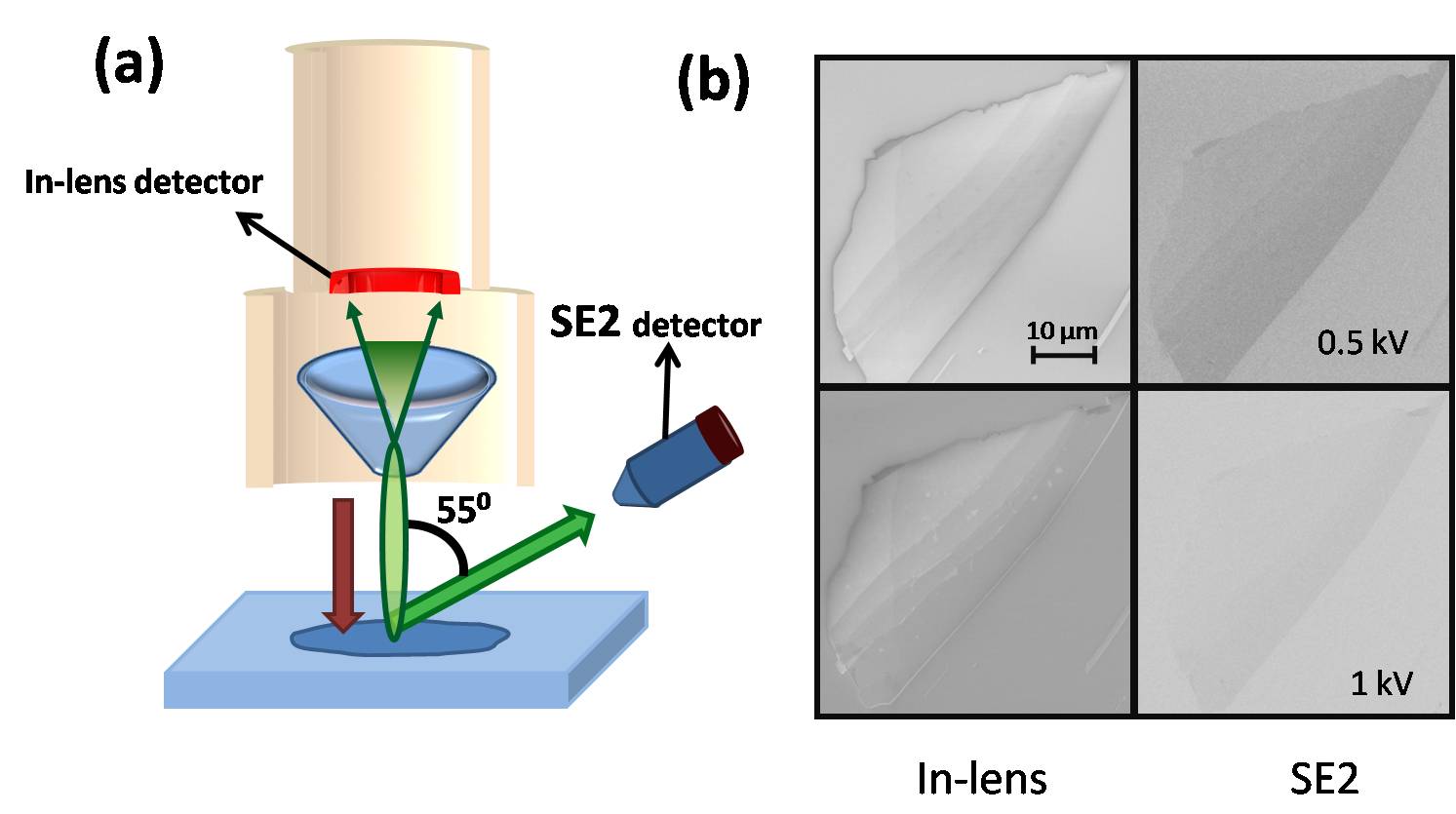}
\end{center}
\caption{ \textbf{(a)} Schematic of secondary electron detection by the in-lens detector placed inside the electron column and the Everhart - Thornley detector (ETD) placed outside the column at an angle of $55^{0}$. \textbf{(b)} The difference in contrast between the images taken by the in-lens detector and the ETD are shown for a typical multilayer graphene sample at two primary electron energies. }
\end{figure}

Recently, attenuation of low-energy electrons in graphene has been exploited
in Auger electron spectroscopy (AES) to determine the number of graphene
layers~\cite{Auger}. The energy relaxation of incident electrons on graphene
occurs over the inelastic mean free path ($\lambda \approx 0.45$~nm below
100~eV), thereby modulating the intensity with varying thickness in the AES
electron maps. Although it is recognized that such a process can yield high
contrast images of multilayer graphene in SEM as well~\cite{apex}, a quantitative
understanding of attenuation of low-energy SE by graphene in an SEM is still lacking.
Experiments using an outer SE detector (often an Everhart-Thornley
detector) generally yield relatively poor contrast, making a quantitative
analysis less reliable. Instead, an in-column SE detector (in-lens) in
a field-emission SEM (FESEM), which generally responds to lower electron
energies ($\sim$ few tens of eV), has been known to produce images of very
sharp contrast for carbon nanotubes on insulating substrates~\cite{voltage contrast,{EBIC}}. Since it is
assumed that the interaction between the SE emitted from the substrate and the nanotube is negligible,
the physical mechanism behind this contrast is under debate, and both
voltage contrast due to differential surface charging~\cite{voltage contrast} and electron beam induced
current (EBIC) processes~\cite{EBIC} have been suggested. Here we show that low-energy SE
imaging using the in-lens detector of a FESEM provides a sharp thickness-dependent
contrast of graphene on insulating substrates over a wide range of primary
electron energies, $E_{p}$. We present a quantitative analysis of the thickness
determination of graphene by taking into account the attenuation of SE of the substrate by the graphene layers.
Due to inherent spatial resolution of SEM (typically $\lesssim 2$~nm at $E_{p}$~$=$~5~kV), this technique can be extremely useful in examining
nanostructures of graphene, such as graphene nanoribbons, and also in
detecting folds, discontinuities and multi-layer growth in large area graphene.

\section{Experimental details}

We have carried out detailed SEM studies on two types of graphene in this work,
namely exfoliated and chemical vapor deposited (CVD) graphene.
Exfoliated graphene flakes were prepared by micromechanical exfoliation of
natural graphite on n$^{++}$ doped silicon substrate covered with 300 nm thick
SiO$_2$ on top. The CVD-graphene was grown by thermal decomposition of methane on
copper at $1000^\circ$C. The graphene was transferred onto the substrate by
coating the copper foil with Poly(methyl methacrylate)~(PMMA 950K), after which the copper was etched off
by placing the foil in an etch solution of Ferric Chloride. The graphene/PMMA
film was then transferred on to the substrate which was then rinsed in acetone,
to dissolve the PMMA leaving graphene on the substrate. The graphene films were characterized by Raman spectroscopy and
AFM to identify the number of layers in various regions.
The SEM images of the graphene films were obtained using an in-lens detector
placed inside the electron column of a Carl Zeiss $\Sigma$IGMA FESEM, and also
an Everhart-Thornley detector (SE2) placed outside the column at an angle of
55$^\circ$ from the column axis. The arrangement is schematically shown in
Fig.~1a. When the primary electron beam is incident on the
sample, low-energy SE and high-energy backscattered electrons (BSE) are
produced from the substrate in the vicinity of the beam exposed region. The SE and BSE are
attracted by the positively biased beam booster of the electron column and are
projected towards the in-lens detector. The BSE are generated from deep within
the substrate and pass through the central aperture of the in-lens detector
without being collected. The in-lens detector collects low-energy SE
efficiently while the SE2 collects high energy SE that are not collected by the
in-lens detector. For preliminary comparison, in Fig.~1b we show a typical
multilayer exfoliated graphene flake containing 2 $-$ 6 layers (see optical
image in Fig.~2a) on SiO$_2$ substrate for two values of $E_p = 0.5$~kV and 1~kV. In both cases
the in-lens image has better contrast, giving graphene greater visibility on SiO$_2$ substrate, than the SE2 image.
This indicates that graphene interacts mainly with low-energy SE, and becomes progressively
transparent to high-energy SE which are collected at SE2.

Given this observation, graphene films were subsequently
imaged with the in-lens detector at varying $E_p$ ranging from 0.5 - 10 kV at a working
distance of 5 mm. The beam current used for the imaging was kept low, in the range of 9 - 16 pA in this voltage regime, in order to reduce the electron beam induced damage to graphene~\cite{{e-beam damage1},{e-beam damage2},{e-beam damage3}}. Strikingly, we find the image contrast to vary strongly
with changing $E_p$ as seen from Fig.2a, and regions having different thicknesses can be
identified from discrete shifts in intensity. The thickness of graphene layers in various
regions of the film has been labeled in the optical micrograph. Note that the
overall contrast as well as that of individual layers is best around $E_p
\simeq 3$~kV, which is similar to that observed in case of carbon nanotubes on SiO$_2$ (around 1 kV)~\cite{voltage contrast,{EBIC}},
thereby indicating that this is due to the nature of the substrate. This
optimum value of $E_p \approx 3$~kV for graphene imaging is an important result.

\section{Results and Discussion}
\vspace{1mm}
\textbf{A. Voltage contrast imaging of graphene layers}
\vspace{1mm}

\begin{figure}[t]
\begin{center}
\includegraphics [width=1\linewidth]{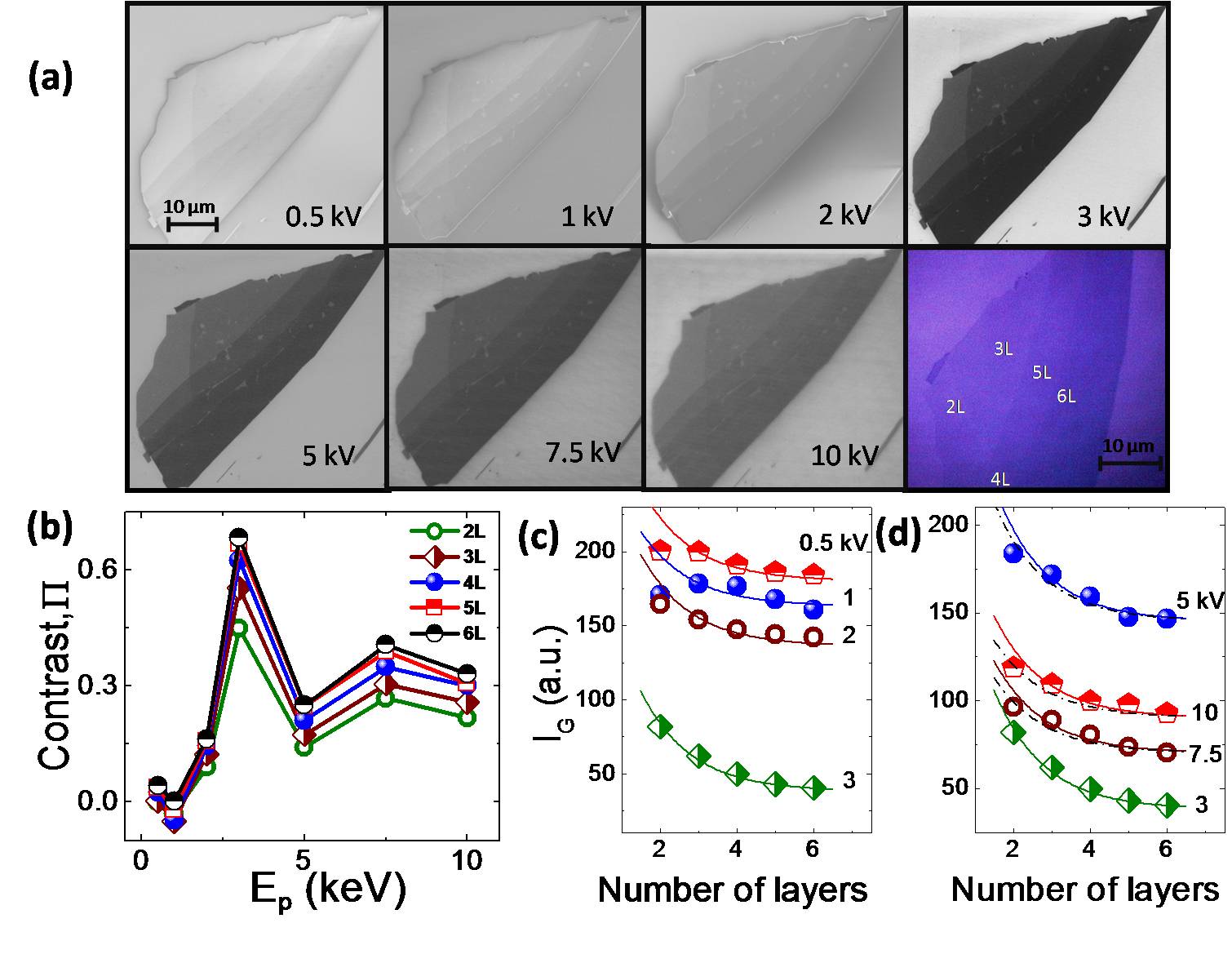}
\end{center}
\caption{ \textbf{(a)} In-lens images for graphene film with 2 - 6 layers  acquired with a scan rate of 7.2 $\mu$m/s for acceleration voltages from 0.5 - 10 kV. The last panel shows an optical micrograph identifying different layers. \textbf{(b)} The contrast parameter (Eq.1) for various layers is plotted as a function of incident beam energy. The contrast between the different graphene layers and SiO$_2$ substrate is maximum at 3kV. \textbf{(c),(d)} The absolute intensity ($I_G$) of SE emission from different graphene layers as a function of number of graphene layers for primary electron energy $\leq$ 3kV \textbf{(c)} and $\geq$ 3kV \textbf{(d)}.  The solid curves denote the fit using Eq.2 with $\alpha = 1$ and the dashed curves represent  $I_{sub}$  multiplied by a fraction which takes into account the negative charging of graphene.
}
\end{figure}

This contrast mechanism can be understood by considering the low ($\lesssim 1$~kV), the intermediate (1kV $\leq E_p \leq$ 3kV) and high
($\gtrsim 5$~kV) regimes of $E_p$ separately. When $E_p \sim 1$~kV, the number of SE emitted from SiO$_2$ surface becomes more than the number of incident electrons which leads to a depletion of electrons from the SiO$_2$ surface. Consequently, the SiO$_2$ surface becomes positively charged which results in a flow of EBIC from graphene to SiO$_2$ surface replenishing that region with electrons and increasing the SE emission from the SiO$_2$ region beneath graphene. This results in negative contrast for $E_p \leq 1$~kV. Further we observed that increasing the beam current while reducing the scan time (or equivalently increasing the scan rate) increases the contrast. In this situation the time taken for SE signal acquisition from a pixel in the substrate is reduced, allowing less time for charge dissipation in the substrate during the SE signal collection, and hence greater pixel-time averaged local potential difference giving greater EBIC and SE signal.

The positive charging of the substrate reduces  in the intermediate kV regime (1.5 - 3 kV), due to which the EBIC from graphene to SiO$_2$ also reduces. Around 3kV, the EBIC is a minimum due to minimal surface charging and the emitted SE are simply attenutated by graphene layers giving rise to a contrast reversal at 3kV. At very high kV ($\geq$ 5kV), a large number of SE are produced in the bulk of SiO$_2$ and Si. But since the SE are produced from depths which are large compared to their escape depth, the number of SE emitted are less leading to negative charging of the substrate. The graphene layers also get charged negatively due to negligible flow of EBIC to SiO$_2$. The thinner regions of graphene which are less metallic, cannot dissipate charge easily and hence a surface negative potential gradient develops in the graphene region leading to greater suppression of SE emission from these regions. These combined effects lead to a reduction in contrast at very high kV.

\vspace{2mm}
\textbf{B. Thickness determination of graphene layers}
\vspace{2mm}

For quantitative analysis, we define the contrast parameter ($\Pi$) as,

\begin{equation}
\label{eq1}
\Pi = \frac{I_{sub}-I_{G}}{I_{sub}+I_{G}}
\end{equation}

\noindent where $I_G$ and $I_{sub}$ are the local intensities at the
graphene region and surrounding substrate (far away from graphene),
respectively, which are proportional to the corresponding number of secondary electrons collected by the in-lens detector. As shown in Fig.~2b, $\Pi$ varies similarly for all layer numbers
and can be as large 0.7 at $E_p \approx 3$~kV. In the discrete shifts in contrast (i.e. $I_G$)
with increasing layer number (Fig. 2a), the thicker regions of graphene always
appear darker than the thinner regions, which implies $I_G$ decreases with
increasing layer number for fixed values of $E_p$, as shown in Fig.~2c ($E_p
\leq 3$~kV) and Fig.~2d ($E_p \geq 3$~kV). We attribute this to attenuation of
the SE within multilayer graphene which can be quantitatively expressed as~\cite{Auger},

\begin{equation}
\label{eq2} I_G (N) = \alpha I_{sub} \exp\left[-\frac{(N-1)d_0 +
d_{vdw}}\lambda\right] + I_0
\end{equation}

\noindent where $N$, $d_0$ ($= 0.335$~nm), and $d_{vdw}$ ($\approx 0.35$~nm)
are the number of layers, inter-layer separation in graphene, and the van der
Waal distance of the graphene from SiO$_2$ substrate, respectively. The factor
$\alpha$ accounts for the negative charging of graphene during electron beam scanning.
All solid lines in Figs.~2c and 2d were obtained with $\alpha = 1$, $\lambda =
0.445$~nm from AES data, and background intensity $I_0$ from the uncovered
parts of SiO$_2$. $I_0$ is the only fitting parameter in Eq.~\ref{eq2}. The overall agreement of Eq.~\ref{eq2} to the observed data confirms
SE attenuation by the graphene layers, and forms a new method to obtain
the thickness of graphene with SEM. We note two extreme cases: (1) For
$E_p \gtrsim 5$~kV, $\alpha \approx 0.9$ gives a marginally better fit (dashed
lines in Fig.~2d), which is due to slight negative charging of graphene which leads to the
suppression of SE emission from the underneath substrate. (2) Secondly, at very
low $E_p \lesssim 1$~kV (Fig.~2c), the SiO$_2$ surface beneath the thinner
regions of graphene, which are less metallic, get significantly positively
charged. This reduces $I_G$ from these regions below the expectation of
Eq.~\ref{eq2}.

\vspace{2mm}
\textbf{C. Contrast reversal in single layer graphene}
\vspace{2mm}

\begin{figure}[t]
\begin{center}
\includegraphics [width=1\linewidth]{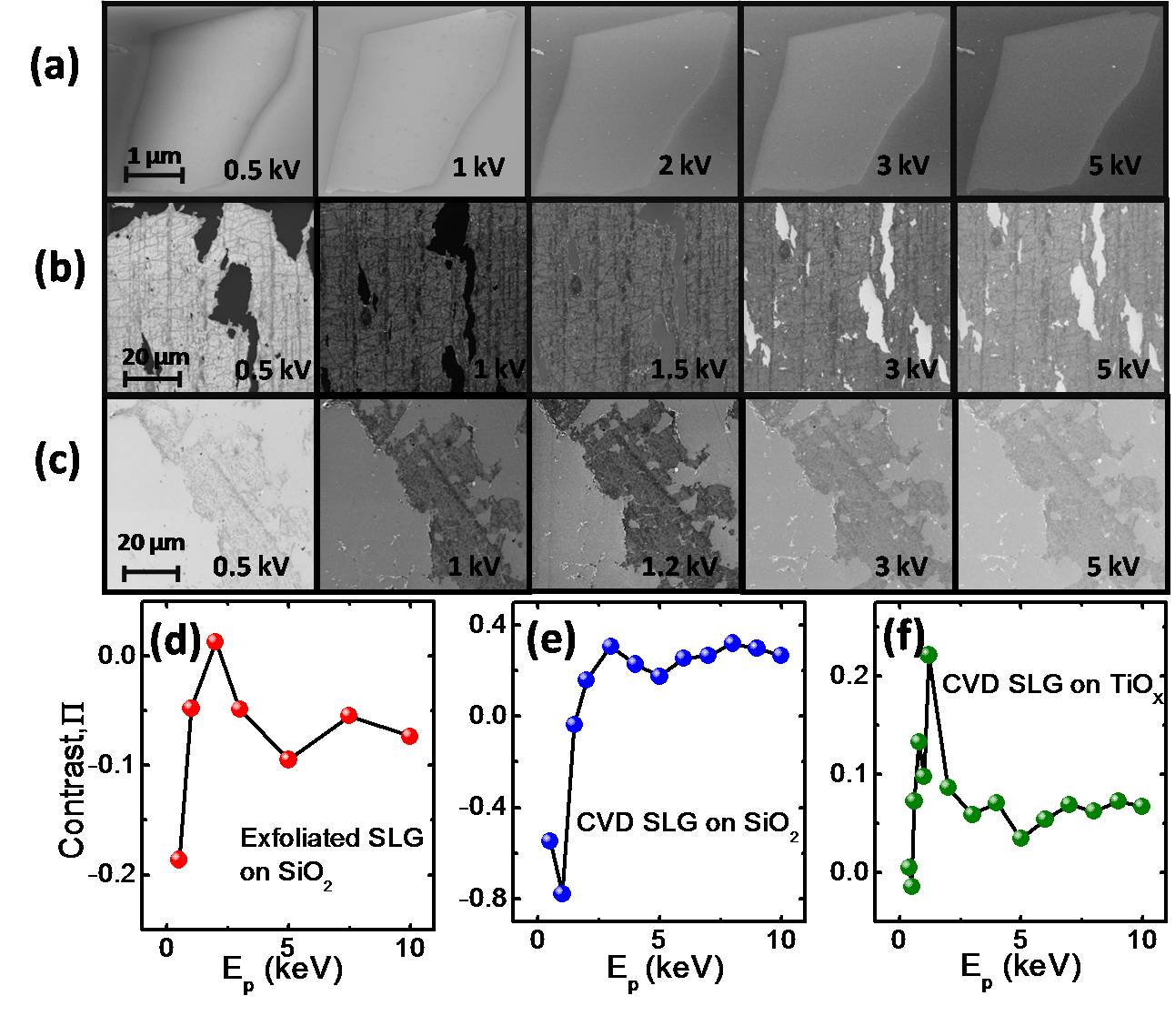}
\end{center}
\caption{\ \textbf{(a)} In-lens images of exfoliated single layer graphene on SiO$_2$/Si substrate for different primary electron energies. \textbf{(b)} In-lens images of CVD grown single layer graphene on SiO$_2$/Si substrate for different primary electron energies showing a sharp contrast reversal below 1kV. Due a small drift at different acceleration voltages, the regions are slightly shifted downwards at higher acceleration voltages. \textbf{(c)} In-lens images of CVD grown single layer graphene on TiO$_x$/Si substrate for different primary energies which do not show a contrast reversal.  All the SEM images in (a) - (c) were acquired in a scan time of 10.4 s. \textbf{(d - f)} Contrast as a function of primary energy for exfoliated SLG on SiO$_2$, CVD grown SLG on SiO$_2$ and CVD grown SLG on TiO$_x$ respectively.}
\end{figure}

The in-lens SEM imaging of single layer graphene (SLG) was found to display
very different behavior with respect to acceleration voltage. Here we have examined three different classes of SLG
devices: (1) Exfoliated SLG on SiO$_2$ substrate, (2) large-area CVD grown
graphene on SiO$_2$ substrate, and (3) CVD graphene on high-$k$ TiO$_x$
substrate. We first focus on the in-lens images of exfoliated and CVD-grown SLG on SiO$_2$
substrate shown in Fig.3a and 3b, respectively. In the exfoliated case (Fig.~3a)
we find that the graphene region appears brighter than the surrounding SiO$_2$
at all $E_p$. This leads to a {\it negative} contrast parameter $\Pi$ (see
Eq.~\ref{eq1}), as opposed to the case of multilayer graphene where $\Pi$ is
positive (Fig.~2b). $\Pi$ is most negative at small $E_p$, becomes nearly zero
at $E_p \sim 2$~kV, but eventually settles to a small negative magnitude ($\sim
-0.07$) at large $E_p$ (Fig.~3d). If we ignore the interaction of SLG with SE,
the brightness of graphene images at low $E_p$ can be explained in a manner
similar to the case of carbon nanotubes~\cite{voltage contrast,{EBIC}}. In this regime ($E_p < 2$~kV)
differential surface charging due to the flow of EBIC from graphene to SiO$_2$ which results in an enhanced SE
emission from the graphene-covered region of the substrate making it appear
brighter. The consistency of this scenario could be readily verified by
replacing SiO$_2$ with atomic layer deposited 350 nm of TiO$_x$ which is a high-$k$ dielectric with dielectric
constant nearly 200. The surface charge is now heavily screened which reduces the EBIC, and hence, CVD-grown graphene on TiO$_x$ substrate
always appears darker at all values
of $E_p$ (Fig.~3c) with a maximum contrast at $E_p \approx 1.2$~kV. At higher $E_p$ ($> 2-3$~kV), SLG becomes negatively charged when scanned by the electron beam. The contrast, whether positive or negative, is determined by the rate at which charge is dissipated from graphene to the substrate, in comparison to the rate at which graphene gets charged by the primary electron beam. A potential equilibrium is achieved more rapidly in the case of exfoliated SLG than the large area CVD graphene since the scan time was kept constant during image acquisition. This explains why $\Pi$ is negative for exfoliated SLG at large $E_p$, and for CVD graphene on SiO$_2$ and TiO$_x$ substrates it saturates at a small positive value
(Fig.~3e). Nonetheless, the reversal of contrast in in-lens SEM as a function of acceleration voltage can serve to
separate the SLG from the multilayer graphene devices.

\begin{figure}[b]
\begin{center}
\includegraphics [width=1\linewidth]{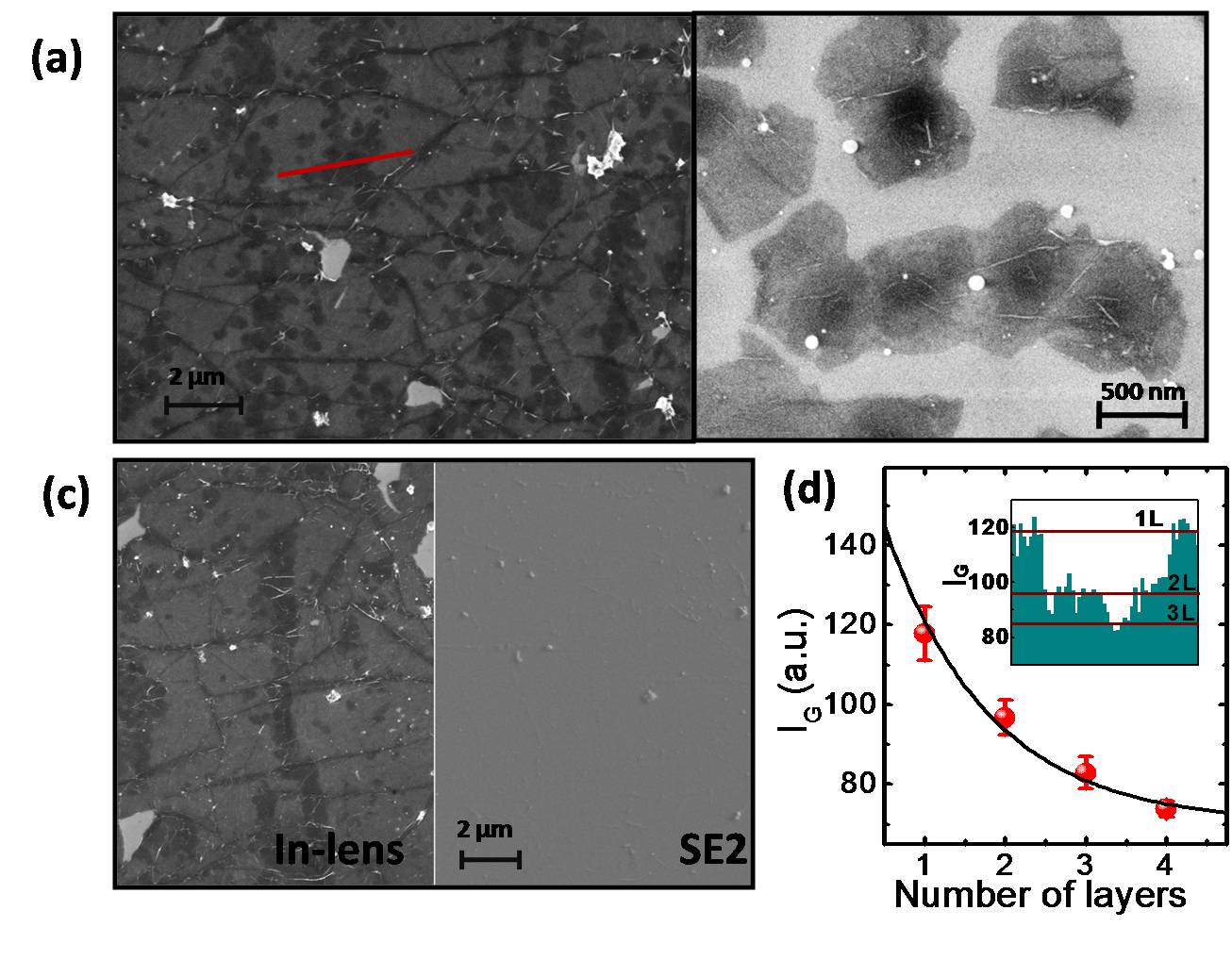}
\end{center}
\caption{ \textbf{(a)} SEM In-lens image of ruptures, folds and regions of multilayer growth in large area graphene grown by chemical vapour deposition on copper and subsequently transferred onto SiO$_2$ substrate. \textbf{(b)} High resolution in-lens image of coalesced sub-$\mu$m sized graphene grains on SiO$_2$ showing excellent contrast. \textbf{(c)} A region of CVD graphene on SiO$_2$ imaged at a primary energy of 3kV using the in-lens detector and the outer detector. \textbf{(d)} The SE intensity of the various multilayer regions in the CVD graphene shown in (a) is fitted by eq.2. The inset shows the SE intensity along a line scan shown in the \textbf{(a)}.}
\end{figure}

\vspace{2mm}
\textbf{D. Applications}
\vspace{2mm}

The high-contrast imaging of graphene layers at low kV can be an effective
nondestructive and quick imaging tool for the structural characterization of voids, ruptures and
folds on graphene at nanometer scale. This is particularly important in large area
graphene, for example those grown by CVD and subsequently transferred onto a different (insulating) substrate.
In Fig.~4a we demonstrate this with an SEM image of CVD-grown (on Cu) graphene transferred on to SiO$_2$
substrate ($E_p = 3$~kV). The wrinkles on CVD graphene observed as branched quasi-linear striations in Fig.4a
possibly represent the texture of the metal on which graphene was grown. The voids and
ruptures which occurred during the transfer process are also easily visible. Additionally, randomly scattered darker patches seem to represent regions of multilayer growth. The multilayered regions are indicated in the intensity profile (inset of Fig.~4d) along the line shown in Fig.~4a. Eq.~\ref{eq2} provides a good fit to
the intensity as a function of layer number (Fig.~4d), albeit with $\alpha
\approx 0.5$ due to significant negative charging of the large area CVD graphene. Fig. 4b shows a high resolution in-lens image of nanometer sized graphene flakes grown by CVD acquired at a very fast scan rate of 450$\mu$m/s, showing excellent contrast even at very large magnification. Finally, in Fig.~4c, a comparison of 3~kV in-lens and SE2 images are shown for a particular region of CVD graphene. It is clear that the SE2 imaging cannot capture the richness
of structure on graphene provided by the in-lens, which is in agreement with the exfoliated case shown in Fig. 1b.

\section{Conclusions}

In summary, we find that SEM imaging using an in-column secondary electron
detector provides a very efficient method of determination of thickness of
graphene films on different insulating substrates. The contrast arises due to
attenuation of secondary electrons from the underlying substrate by multilayer graphene, characterized by an inelastic mean free path. Differential charging of the
substrate due to EBIC affects the contrast at low primary electron energies, and causes a
``contrast reversal'' phenomenon in single layer graphene on SiO$_2$. We have
also demonstrated that this technique can be exploited for structural
characterization of graphene down to nanometer length scales.

\vspace{2mm}
\textbf{Acknowledgment}
\vspace{2mm}

We acknowledge the Department of Science and Technology (DST) for a funded project. S.R. acknowledges support under Grant No. SR/S2/CMP-02/2007. V.K. and A.N.P. thank CSIR for financial support.


\begin{thebibliography}{30}


\bibitem{bilayer} J. B. Oostinga, H. B. Heersche, X. Liu, A. F. Morpurgo and  L. M. K. Vandersypen, Nature Mater. \textbf{7}, 151 (2007).

\bibitem{trilayer}  M. F. Craciun, S. Russo, M. Yamamoto, J. B. Oostinga, A. F. Morpurgo and S. Tarucha, Nature Nanotech. \textbf{4}, 383 (2009).

\bibitem{thermopower} S. Ghosh, W. Bao, D. L. Nika, S. Subrina, E. P. Pokatilov, C. N. Lau and A. A. Balandin, Nat. Mater. \textbf{9}, 555 (2010).

\bibitem{mechanical} I. W. Frank, D. M. Tanenbaum, A. M. van der Zande and P. L. McEuen, J. Vac. Sci. Technol. B \textbf{25}, 2558 (2007).

\bibitem{optical_1} P. Blake, K. S. Novoselov, A. H. Castro Neto, D. Jiang, R. Yang, T. J. Booth, A. K. Geim and E. W. Hill, Appl. Phys. Lett. \textbf{91}, 063124 (2007).

\bibitem{McCann_bilayer} E. McCann, Phys. Rev. B \textbf{74}, 161403(R) (2006).

\bibitem{atin APL}  A. N. Pal and A. Ghosh, Appl. Phys. Lett. \textbf{95}, 082105 (2009).

\bibitem{atin PRL} A. N. Pal and A. Ghosh, Phys. Rev. Lett. \textbf{102}, 126805 (2009).

\bibitem{noise arxive} A. N. Pal, S. Ghatak, V. Kochat, E. S. Sneha, B. S. Arjun, S. Raghavan and A. Ghosh,  \emph{e-print arXiv:condmat}/1009.5832v2 (2010).

\bibitem{SiC} C. Berger, Z. Song, X. Li, X. Wu, N. Brown, C. Naud, D. Mayou, T. Li, J. Hass, A. N. Marchenkov, E. H. Conrad, P. N. First and W. A. de Heer, Science \textbf{312}, 1191 (2006).

\bibitem{CVD_science} X. Li, W. Cai, J. An, S. Kim, J. Nah, D. Yang, R. Piner, A. Velamakanni, I. Jung, E. Tutuc, S. K. Banerjee, L. Colombo and R. S. Ruoff, Science \textbf{324}, 1312 (2009).

\bibitem{CVD_nanoletters} A. Reina, X. Jia, J. Ho, D. Nezich, H. Son, V. Bulovic, M. S. Dresselhaus and J. Kong, Nano Lett. \textbf{9}, 30 (2009).

\bibitem{optical_2} P. E. Gaskell, H. S. Skulason, C. Rodenchuk and T. Szkopek, Appl. Phys. Lett. \textbf{94}, 143101 (2009).

\bibitem{AFM} P. Nemes-Incze, Z. Osvátha, K. Kamarásb and L. P. Biróa, Carbon \textbf{46}, 1435 (2008).

\bibitem{Raman_1} A. C. Ferrari, J. C. Meyer, V. Scardaci, C. Casiraghi, M. Lazzeri, F. Mauri, S. Piscanec, D. Jiang, K. S. Novoselov, S. Roth and A. K. Geim, Phys. Rev. Lett. \textbf{97}, 187401 (2006).

\bibitem{Raman_2} A. Das, S. Pisana, B. Chakraborty, S. Piscanec, S. K. Saha, U. V. Waghmare, K. S. Novoselov, H. R. Krishnamurthy, A. K. Geim, A. C. Ferrari and A. K. Sood, Nature Nanotech. \textbf{3}, 210 (2008).

\bibitem{Raman_referee} N. Camara, J-R. Huntzinger, G. Rius, A. Tiberj, N. Mestres, F. Pérez-Murano,  P. Godignon and J. Camassel, Phys. Rev. B \textbf{80}, 125410 (2009).

\bibitem{contrast spectroscopy} Z. H. Ni, H. M. Wang, J. Kasim, H. M. Fan, T. Yu, Y. H. Wu, Y. P. Feng and Z. X. Shen, Nano Lett. \textbf{7}, 2758 (2007).

\bibitem{Rayleigh spectroscopy} C. Casiraghi, A. Hartschuh, E. Lidorikis, H. Qian, H. Harutyunyan, T. Gokus, K. S. Novoselov and A. C. Ferrari, Nano Lett. \textbf{7}, 2711 (2007).

\bibitem{LEED} C. Riedl, A. A. Zakharov and U. Starke, Appl. Phys. Lett. \textbf{93}, 033106 (2008).

\bibitem{Auger} M. Xu, D. Fujita, J. Gao and N. Hanagata, ACS Nano \textbf{4}, 2937 (2010).

\bibitem{apex} H. Hiura, H. Miyazaki and K. Tsukagoshi, Appl. Phys. Exp. \textbf{3}, 095101 (2010).

\bibitem{voltage contrast} T. Brintlinger, Y-F. Chen, T. Dürkop, E. Cobas, M. S. Fuhrer,  J. D. Barry and J. Melngailis, Appl. Phys. Lett. \textbf{81}, 2454 (2002).

\bibitem{EBIC} Y. Homma, S. Suzuki, Y. Kobayashi, M. Nagase and D. Takagi, Appl. Phys. Lett. \textbf{84}, 1750 (2004).

\bibitem{e-beam damage1} G. Rius, A. Verdaguer, F. A. Chaves, I. Martín, P. Godignon, E. Lora-Tamayo, D. Jiménez and F. Pérez-Murano, Microelectron. Eng. \textbf{85}, 1413 (2008).

\bibitem{e-beam damage2} D. Teweldebrhan and A. A. Balandin, Appl. Phys. Lett. \textbf{94}, 013101 (2009).

\bibitem{e-beam damage3} I. Childres, L. A. Jauregui, M. Foxe, J. Tian, R. Jalilian, I. Jovanovic and Y. P. Chen, Appl. Phys. Lett. \textbf{97}, 173109 (2010).


\end{thebibliography}
\end{document}